\documentclass[prd,preprint,nofootinbib,showpacs]{revtex4}
\usepackage{graphicx}
\usepackage{bm}

\newcommand{\bi}{\bar{\imath}}
\newcommand{\bj}{\bar{\jmath}}
\newcommand{\bk}{{\bar k}}
\newcommand{\bl}{{\bar l}}
\newcommand{\be}{\begin{equation}}
\newcommand{\ee}{\end{equation}}
\newcommand{\bea}{\begin{eqnarray}}
\newcommand{\eea}{\end{eqnarray}}
\newcommand{\bz}{\bar{z}}

\begin{document}

\bigskip
\begin{flushright}
NSF-PHY97-22022\\
hep-th/0209200
\end{flushright}
\bigskip
\bigskip

\title{MSSM parameters from supergravity backgrounds}

\author{Mariana Gra\~na}
\email{mariana@physics.ucsb.edu}
\affiliation{Department of Physics\\ University of California\\  
Santa Barbara, CA 93106}
\bigskip
\begin{abstract}
\bigskip
We find the parameters of the MSSM in terms of bulk supergravity fields for the D-brane model of Berenstein, Jejjala and Leigh (hep-ph/0105042). The model 
consists of a D3-brane at the singularity of a non-abelian orbifold $\Delta_{27}$, which gives the particles of the Supersymmetric Standard Model. We compute 
the action for the D-brane  fields in the presence of both supersymmetric and supersymmetry breaking background fluxes. We get quark, lepton, 
gaugino, Higgsino, scalar partners and Higgs boson masses  as well as soft trilinear 
couplings as functions of the background fields. This work develops a framework for connecting MSSM phenomenology to  
brane compactifications.    
\end{abstract}

\date{\today}
\bigskip
\bigskip
\maketitle
\newpage

\section{Introduction}

The Standard Model of Particle Physics has passed stringent quantitative experimental tests. 
This means that if string theory is the underlying theory, it should contain as a low energy limit the Standard Model.
The viewpoint on how to embed the Standard Model in string theory has changed over the past five years.
The heterotic string no longer has a monopoly, yielding its place to (heterotic) M-theory, type I and type II
strings. 

Semirealistic models have been constructed using different types of compactifications (see 
\cite{SM95} and references therein). Another approach, one of the most popular nowadays, arises after the realization that 
gauge fields in type I theory live on the world-volume of a Dp-brane (see \cite{TASI} for review), while gravity fields
are realized as closed strings living in the bulk. The hierarchy between the Planck scale and the weak scale is
explained by large extra dimensions \cite{Dimopoulos} or a warped space \cite{RS,GKP, DWG}. The string scale  
is no longer tied to the Planck scale, and can be as low as the experimental bound of 1 TeV.

By combining groups of rotated branes and antibranes, extended in 4 dimensions and wrapped in the others, with or without orbifolds, 
several models with the matter content of the Standard Model  have been built
\cite{SM>95}.  Since the gauge group on a set of $N$ D-branes is $SU(N)$, these models contain usually 
sets of branes in groups of 2 or 3 to give the Standard Model gauge group. A model first proposed, among various others,  by Aldazabal, Iba\~nez,  Quevedo and
Uranga \cite{AIQU}, and taken up by Berenstein, Jejjala and Leigh a year ago \cite{Berenstein}, stands
out for its simplicity: it contains just one D3-brane. The way to get the Standard Model is by orbifolding the 6-dimensional
space by a non-abelian subgroup of $SU(3)$ ($\Delta_{27}$), thus preserving ${\mathcal N}=1$ supersymmetry. This model realizes all the particles of the 
Supersymmetric Standard Model (SSM), with six Higgs doublets and right handed neutrinos. 

D-brane fields interact with  bulk fields, modifying their masses and couplings. The precise result of this interaction 
for the models built so far has not been determined yet, to our knowledge. In this paper, we will get the Lagrangian for 
the Standard Model particles in the presence of background fluxes, for the model built in \cite{Berenstein}. In consequence,   
we will get all the parameters of the softly broken SSM in terms of the (supersymmetry breaking) background fluxes, thus building the 
bridge between MSSM phenomenology  and string configurations in background fluxes.

The paper is organized as follows. In Section II, we present the model of Berenstein, Jejjala and Leigh that leads 
to the particle content of the SSM. In Section III we find the background that survives the orbifold projection, preserves
Lorentz invariance on the world-volume and solves the equations of motion,
as a power series in the distance to the orbifold point.  To get the Lagrangian for a D3-brane in the $\Delta_{27}$ orbifold, we start with the non-abelian 
Dirac-Born-Infeld and Wess-Zumino bosonic actions in the background found in Section II. For the fermionic action, we will use the abelian one 
for a D3-brane computed in \cite{Grana}, and find any non-abelian additional terms with the help of the non-abelian 
action for D0-branes \cite{Taylor2}. All this is done in Section III. In Section IV, we find the parameters of the MSSM in terms of background fluxes, and we state our conclusions in Section V. Appendix A shows the details of the non-abelian group by which the orthogonal directions are orbifolded,
and explains how to get the matter content for a D-brane at the orbifold point.

\section{The Model}

If we orbifold the space orthogonal to a D3-brane by a discrete subgroup 
$\Gamma$ of $SU(3)$, we get on the brane an ${\mathcal N}=1$, $\Pi_{a=1}^r 
U(N_a)$ gauge theory, where $N_a$ are the dimensions of the irreducible 
representations of $\Gamma$, with chiral multiplets transforming as 
$(N_a,\overline{N}_b)$ \cite{Lawrence}. It was shown in \cite{Berenstein} that 
when $\Gamma$ is the non-abelian group $\Delta_{27}$, the matter obtained 
from keeping states invariant under the orbifold projection resembles very 
closely  that of the standard model, with three quark and lepton 
generations, neutrino singlets and six Higgs doublets.

The miracle works as follows \cite{Greene, Muto, Hanany}: to find states 
invariant under the orbifold projection, we have to consider the D3-brane 
and all of its images, making a total of $|\Gamma|$ ($|\Gamma|$ is the order 
-number of elements- of the group) D3-branes. In the case of the group 
$\Delta_{27}$, there are $27$ D3-branes to start with, and the original 
world-volume theory is an ${\mathcal N}=1$, $U(27)$ gauge theory, with a 
$U(27)$ vector multiplet $V$, made out of the gauge fields and the gauginos 
$\lambda$,  and three chiral multiplets $\Phi^i$ composed by the complex 
scalars $\phi^i$, identified with the orthogonal complex coordinates $z^i$, 
and three Weyl fermions $\psi^i$, all of them $N\times N$ matrices, 
transforming in the adjoint representation. Projecting onto $\Gamma$ 
invariant states means satisfying
\bea
R_{reg} V R_{reg}^{-1}= V \label{Vproj}\\
(R_3)_{ij} R_{reg} \Phi^j R_{reg}^{-1}=\Phi^i \label{phiproj}
\eea
where $R_{reg}$ is the $N\times N$ regular representation acting on the 
Chan-Paton index, and $R_3$ is the 3-dimensional defining representation 
that acts on the space-time index $i$. Any representation can be decomposed 
into a sum of irreducible representations $R^a$. For the regular 
representation, the decompositions works as follows
\be
R_{reg}=\oplus_{a=1}^r N_a R^a
\ee
where $N_a=dim R^a$, i.e. each irreducible representation $R^a$ occurs $N_a$ 
times in the regular representation. The regular representation then has the 
form
\bea
R_{reg}= \pmatrix{R^1 \otimes 1_{N_1} & 0 & ... & 0 \cr
0 & R^2 \otimes 1_{N_2} & ... & 0 \cr
\vdots& \vdots & \ddots & \vdots \cr
0 & 0 & ... & R^r \otimes 1_{N_r} } \cr
\eea
where $R^a \otimes 1_{N_a}$ is the $N_a \times N_a$ matrix with $N_a$ copies 
of $R^a$
\bea
R^a \otimes 1_{N_a} =  \pmatrix{R^a & 0 & ... & 0 \cr
0 & R^a & ... & 0 \cr
\vdots& \vdots & \ddots & \vdots \cr
0 & 0 & ... & R^a  } \cr
\eea

From Eq.(\ref{Vproj}) we get that the gauge symmetry is
\be
\prod_{a=1}^r U(N_a)
\ee
The group $\Delta _{27}$ has two 3-dimensional and nine 1-dimensional 
representations. Thus the gauge theory on the D3-brane is a $U(3)^2\times 
U(1)^9$. The first of these $U(3)$'s will be identified with color, and the 
second one, combined with some of the $U(1)$'s, and after breaking part of 
the symmetry, with $SU(2)\times U(1)$ of the electroweak interactions. At the 
weak scale only the hypercharge plus two other $U(1)$ symmetries will 
survive.
The condition (\ref{phiproj}) tells us the number of chiral fields 
$n^3_{ab}$ transforming in the $(N_a, \overline{N}_b)$, by decomposing the 
product of the defining and each irreducible representation into irreducible 
representations in the following way:
\be
R_3 \otimes R^a = \oplus_{b=1}^r n^3_{ab} R^b.
\label{couplings}
\ee
It is shown in the Appendix how to get the numbers $n^3_{ab}$ from the table 
of characters of a group. In the case of the group $\Delta_{27}$, we get 3 
chiral multiplets transforming as ${\cal Q}_i=(3,\overline{3},0)$ of $U(3)_c 
\times U(3)_w \times U(1)^a$, and one multiplet for each $U(1)$ group, 
transforming as ${\cal L}_a=(1,3,-_a)$ and $\overline{{\cal 
Q}}_a=(\bar{3},1,+_a)$, using the notation of \cite{Berenstein}. Gauge couplings
are given by
\be
\tau_a=\frac{N_a \tau}{|\Gamma|}
\ee

The quiver 
of the theory is shown in Figure 1.

\begin{figure}
\includegraphics{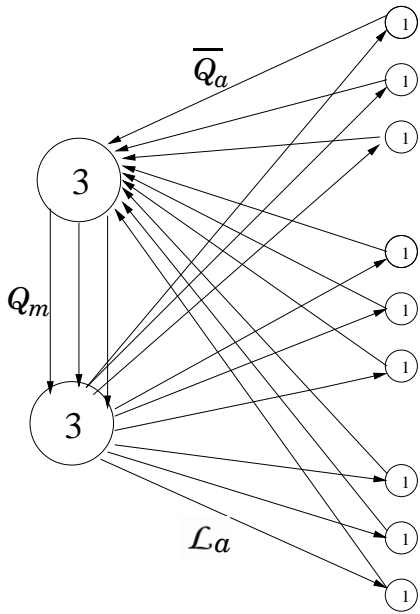}
\caption{\label{f:quiver} Quiver diagram for the group $\Delta_{27}$ }
\end{figure}

As a first step, Berenstein, Jejjala and Leigh add Fayet-Iliopoulos 
parameters for three of the $U(1)$'s, enabling a supersymmetry preserving 
vacuum expectation value for three scalars in the chiral multiplet ${\cal 
L}_a$. This breaks $U(3)_w \times U(1)^3$ into $SU(2)_w \times U(1)_0$. Then 
all fields transforming in the fundamental representation of $U(3)_w$ get 
split into an $SU(2)$ doublet and an $SU(2)$ singlet. Then, we can find all 
the supersymmetric Standard Model particles in the spectrum of this theory 
as follows:
\bea \label{matter}
{\cal Q}_m &\rightarrow Q_m \ , \ q_m \qquad \qquad &\bar{{\cal Q}}_{1,m} 
\equiv \bar{q}_m \nonumber \\
{\cal L}_{1,m} &\rightarrow L_m \ , \ g_m \qquad \qquad &\bar{{\cal 
Q}}_{2,m} \equiv \bar{u}_m \\
{\cal L}_{2,m} &\rightarrow H_m \ , \ \bar{e}_m \qquad \qquad &\bar{{\cal 
Q}}_{3,m} \equiv \bar{d}_m \nonumber\\
{\cal L}_{3,m} &\rightarrow {\overline H}_m \ , \ \bar{\nu}_m \qquad \qquad & \nonumber
\eea
where the fields on the left are those that transform as a $3$ of $U(3)_w$, 
and we give their decomposition into $SU(2)_w$ doublets and singlets. The 
index $a$ has been decomposed into a pair $l,m$, and the fields charged 
under the three $U(1)$'s of the form ``$1,m$'' (called $g_m$ in 
Eq.(\ref{matter})) are those that get VEVs.

As we can see, we get all of the fields of the Standard Model. The index $m$ 
is a generation index, so we get three quark generations and three lepton 
generations, with  both $SU(2)_w$ doublets and singlets (we get a neutrino 
singlet), and six Higgs doublet fields. Besides, there are fields not present 
in the Standard Model, namely $g_m$, which get vevs that break one of the 
$U(3)$'s into $SU(2)_w$, and those called $q_m$ and $\bar{q}_m$, which get 
mass from the superpotential coming from ${\mathcal N}$=4, and can be 
integrated out.

It is remarkable that with just one D3-brane we can get the matter content 
of the Standard Model.

\section{Background}

In this Section we are going to immerse the D3-brane in background fluxes.
But not any flux survives the orbifold projection, so we will find first those
that do survive and also
solve the equations of motion. We will find them as a series 
expansion in $z^i$, the complex distance to the orbifold singularity, up to 
second order.

To preserve Lorentz invariance on the D3-brane, we want the metric to be of the warped form, so in its longitudinal part there can be at most a warp factor multiplying the Minkowski metric. For the remaining directions, we can have any metric component ($dz^i dz^j$, $d\bz^i d\bz^j$ and $dz^i d\bz^j$), as long as they give a metric invariant under the orbifold operations. The rest of the fields (axion-dilaton, 5- and 3-form fluxes) can only be a function of $z^i$ and $\bz^i$. We will divide the 3-form flux into pieces according to the number of holomorphic and antiholomorphic indexes: (3,0); (0,3); (1,2) and (2,1). It will be convenient to express the (1,2) and (2,1) pieces in
terms of a symmetric and an antisymmetric 2-tensor, in the way shown below.     

The only possible expansion of these fields compatible with the orbifold are \footnote{Basically, a scalar field is invariant under the orbifold operations
(\ref{elements}) if it is proportional to the radius $z^i \bz^i$; a vector $V_i$ has to be proportional to $\bz^i$, a tensor $V_{ij}$ to either $\bz^i \bz^j$ or $\epsilon_{ijk} z^k$, and so on.} 
\bea \label{fields}
ds^2&=&\left(1+k_2z^i \bar{z}^i\right) \eta_{\mu \nu} dx^{\mu} dx^{\nu}+ \left((1+h_2 z^k \bar{z}^k) \delta_{ij}+ l_2 \bar{z}^i z^j\right) dz^i d\bar{z}^j \nonumber\\
&& + g_2 \bz^i \bz^j dz^i dz^j + g_2^* z^i z^j d\bz^i d\bz^j\nonumber\\ 
G_{123}& =&G_0 + G_2 z^i \bz^i , \qquad G_{\bar{1}\bar{2}\bar{3}} =G'_0 + G'_2 z^i \bz^i\nonumber \\
S_{ij}&\equiv& \frac{1}{2}\left(\epsilon_{ikl}G_{{\bar k} {\bar l} j}+\epsilon_{jkl}G_{{\bar k}{\bar l}i}\right)= s_2 \bz^i\bz^j\qquad  \nonumber\\
S_{\bi \bj}&\equiv&\frac{1}{2}\left(\epsilon_{\bi {\bar k} {\bar l}} 
G_{kl\bj}+\epsilon_{\bj \bk \bl}G_{kl\bi}\right)= s'_2 z^i z^j \nonumber\\
A_{ij}&\equiv& \frac{1}{2}\left(\epsilon_{ikl}G_{\bar{k}\bar{l}j}-\epsilon_{jkl}G_{\bar{k}\bar{l}i}\right)= a_1  \epsilon_{ijk}z^k \nonumber\\
 A_{\bi \bj}&\equiv&\frac{1}{2}\left(\epsilon_{\bi {\bar k} {\bar l}} 
G_{kl\bj}-\epsilon_{\bj {\bar k} {\bar l}}G_{kl\bi}\right)= - a_1 \epsilon_{\bi \bj {\bar k}}\bar{z}^k\nonumber\\
\tau&=&\tau_0+\tau_2 z^i \bar{z}^i, \qquad  F_{i \bi j \bj k}= F_1 \epsilon_{i \bi j \bj k \bar{k} } \bz^k 
\eea
where $k_2,h_2,l_2$ and $F_1$ are real constants, and $g_2,  s_1, s'_1, a_1, \tau_0, \tau_2, G_0,G_2$ are complex numbers\footnote{The constant $a_1$ in the first order $A_{ij}$ and $A_{\bi \bj}$ is the same since they both come from the second order (1,1) piece of $B_{(2)}$ and $C_{(2)}$.} and $\epsilon_{i \bi j \bj k 
\bar{k}}=i$  

The fields in Eq.(\ref{fields}), with any value of the constants, satisfy the orbifold projection but not necessarily the equations of motion. 
In general, the equations of motion give relationships between the constants, as we will show.

The field equations for type IIB supergravity are 
\begin{eqnarray} \label{EOM}
\bigtriangledown^m \left(\frac{i}{2} e^{\phi} \partial_m \tau \right)&-& \frac{1}{2}e^{2\phi} \partial^m C \partial_m \tau =\frac{g_s}{24}G_{mnp}G^{mnp}\, \nonumber\\
\bigtriangledown^mG_{mnp} &+& \frac{i}{2} e^{\phi} \partial^m C G_{mnp} =- \frac{i}{2} e^{\phi} \partial^m\tau G_{mnp}^\ast-\frac{g_s}{6}iF_{mnpqr}G^{mqr} \, , \nonumber \\
R_{mn}&=& \frac{1}{4} e^{2\phi} \partial_m\tau \partial_n\tau^\ast +  \frac{1}{4} e^{2\phi} \partial_m\tau^\ast \partial_n\tau +\frac{g_s^2}{96}F_{mqrst}{F_n}^{qrst} 
\\
&&+\frac{g_s}{8}\left({G_m}^{pq}G_{npq}^\ast +{G_n}^{pq}G_{mpq}^\ast-
\frac{1}{6}g_{mn}G_{pqr}^\ast G^{pqr}\right)\, . \nonumber 
\end{eqnarray}
The Bianchi identities are
\begin{equation}
\partial_{[m}G_{nkl]}+ \frac{i}{2} e^{\phi} \partial_{[m}C G_{nkl]}=\frac{i}{2} e^{\phi} \partial_{[m}\tau G^\ast_{nkl]}, \qquad
\partial_{[m}F_{nklpq]}=\frac{5}{48}iG_{[mnk}G^\ast_{lpq]}
\label{Bianchi}
\end{equation}

Inserting the fields (\ref{fields}) in the equations of motion (\ref{EOM}) and Bianchi identities (\ref{Bianchi}), we get relationships between the constants appearing in the fields. Solving them requires
\bea
a_1&=&0
\label{A} \\
12 F_1 &=&|G_0|^2-|G'_0|^2 \label{F1det}\\
\tau_2&=&-\frac{i}{3} G_0 G'_0 \label{tau2det}\\
24 k_2 &=& g_s\left(|G_0|^2+|G'_0|^2\right) \label{k2det}\\
k_2&=& \frac{1}{5} \left(8 Re g_2 + l_2 - 5 l_2 \right) \label{rel0}\\
F_1 G'_0 g_s &=&-\frac{i}{2}g_s \left(Re \tau_2 G'_0 + \tau_2 (G'_0)^\ast\right)+8g_2 G'_0- 4s_2-G'_2 \label{rel1}\\
F_1 G_0 g_s &=&-\frac{i}{2}g_s \left(Re \tau_2 G_0 + \tau_2 (G_0)^\ast\right)+8g^\ast_2 G_0 -4s'_2-G_2 \label{rel2}\\
0&=&\frac{i}{2}g_s \left(Re \tau_2 G'_0 - \tau_2 (G'_0)^\ast\right)+2s_2+G'_2 \label{rel3}\\
0&=&\frac{i}{2}g_s \left(Re \tau_2 G_0 - \tau_2 (G_0)^\ast\right)+2s'_2+G_2\label{rel4}\
\eea
where $g_s=1/Im \tau_0$.  

Any configuration of the form (\ref{fields}) satisfying Eqs.(\ref{A}-\ref{rel4}) is a consistent background. 

At zeroth order, the free parameters are the axion-dilaton $\tau_0$, and the (3,0) and (0,3) constant 3-form flux pieces $G_0$ and $G'_0$.  
$G_0$ and $G'_0$ determine the first order parameter $F_1$ in the 5-form flux, as can be seen in 
Eq.(\ref{F1det}), and the second order $\tau_2$ and $k_2$ in the  dilaton and longitudinal metric, respectively, as shown in Eqs.(\ref{tau2det}) and (\ref{k2det}). The remaining six second order parameters ($G_2, g_2, s_2, s'_2, l_2$ and $h_2$) are constrained by five equations, Eqs.(\ref{rel0})-(\ref{rel4}) so there is some freedom to play with. 

The four supersymmetries preserved by the orbifold are the same as those studied in \cite{GP, Gubser}, where it was shown that the (3,0); (0,3) and symmetric (1,2) pieces of the 3-form flux, as well as a non-vanishing antiholomorphic derivative of the dilaton, break supersymmetry. Then, a vacuum expectation value for  $G_0$, $G'_0$, $G_2$, $G'_2$, $s_2$ and $\tau_2$ break supersymmetry. As we will see, $G_0$ will generate gaugino masses and soft trilinear A-terms. $s_2$ will appear in lepton masses and in the $\mu$-term, and $F_1, k_2$ and $\tau_2$ will combine to give masses to the scalar partners.    

\section{Action}

We wish to find the non-abelian action corresponding to 27 D3-branes at the orbifold point, in the background found in the previous Section. The bosonic part of action is known, it is 
the non-abelian Born Infeld and Wess Zumino actions \cite{Myers, Taylor1}, that we can trust up to terms of order $F^4$. This is enough 
for us, since we just want the $F$-independent terms.
In the case of the fermionic action, the non-abelian version of it is not known yet. For a flat background, we know the action will be
that of ${\mathcal N}=1$, $U(27)$ super-Yang Mills with the superpotential 
 $W=Tr \left([\Xi^i,\Xi^j] \Xi^k\right)$ ($\Xi^i$ is the chiral superfield containing $\Phi^i$ and $\psi^i$). When nontrivial fluxes are present, the brane-background fermionic couplings for a D3-brane have been found in \cite{Grana} for just one D3-brane, and these will appear in the non-abelian action with a trace over the gauge indexes. Finally,  we will check that there are not any additional couplings than those already found, using the T-dual of the non-abelian action linear in the background for $N$ D0-branes found from matrix theory in \cite{Taylor2}.

Let us start with the bosonic action (we use the notation of \cite{Myers}):
\bea
S=&-&\mu_3 \int d^4 \zeta\, Tr\left(e^{-\phi} \sqrt{-det\left(P\left[E_{\mu\nu}+E_{\mu i}\,(Q^{-1}-\delta)^{ij}E_{j\nu}\right]+2\pi \alpha' F_{\mu\nu}\right) det (Q^i_j)}\right) \nonumber\\
&+& \mu_3 \int Tr\left(P\left[e^{i2\pi \alpha' i_{\Phi} i_{\Phi}} \sum C_{(n)} e^B\right] e^{2 \pi \alpha' F}\right)
\eea
where $P$ is the pull-back of the spacetime fields into the world-volume, defined
\be
P\left[E\right]_{\mu \nu}= E_{\mu \nu}+2\pi \alpha'E_{i(\mu}D_{\nu)}\Phi^i+4 \pi^2 \alpha'^2 E_{ij} D_\mu \Phi^i D_\nu \Phi^j, \qquad D_\mu\Phi^i=\partial_\mu \Phi^i + i [A_\mu, \Phi^i]
\ee
the scalars $\Phi^i$ are the orthogonal coordinates $z^i$ rescaled by a factor $2\pi \alpha'$ as: $\Phi^i=\frac{z^i}{2\pi \alpha'}$, the tensor $E$ and $Q$ are defined
\be
E_{\mu\nu}\equiv G_{\mu\nu}+B_{\mu\nu}, \qquad Q^i_j\equiv \delta^i_j+i 2\pi \alpha' [\Phi^i,\Phi^k] E_{kj}
\ee
and $i_{\Phi}$ denotes the interior product by $\Phi$
\be
i_{\Phi}i_{\Phi} C_{(2)}= \Phi^j \Phi^i C_{ij}.
\ee

Inserting the background (\ref{fields}), and 
expanding the square root in the Dirac-Born-Infeld action, we get, up to terms of dimension four,
\bea \label{actionbos}
{\mathcal L}=\frac{1}{g_s}\, Tr \left( \frac{1}{2} \partial_\mu \Phi^i \partial^\mu \Phi^{\bi}+ (-2k_2-g_s F_1- g_s Im\tau_2)\,\Phi^i \Phi^{\bi} \right. \nonumber\\
 - \frac{1}{3} G_0 \epsilon_{ijk}\Phi^i\Phi^j\Phi^k - \frac{1}{3} G'_{0}\epsilon_{\bi\bj\bar{k}}\Phi^{\bi}
\Phi^{\bj}\Phi^{\bar{k}} 
\left. - \frac{1}{2}[\Phi^j, \Phi^{(i}][\Phi^{\bi)}, \Phi^{\bj}] \right)
\eea 

For the fermionic action, we first recall the abelian one obtained from the kappa-symmetric action \cite{Grana}:
\be
{\mathcal L}= - \frac{1}{2} \bar{\psi}^{\bi} \Gamma^\mu D_\mu \psi^i -\frac{1}{2} \bar{\lambda} \Gamma^\mu D_\mu \lambda
+\frac{1}{48}\psi^i\psi^j S_{ij}+\frac{1}{48}\lambda \lambda G_{123}+\frac{1}{96}\epsilon_{ijk}\psi^i\lambda A_{\bj{\bar k}}
\label{actionabel}
\ee
where $\psi^i$ are the superpartners of $\Phi^i$, and $\lambda$ is the gaugino

The non-abelian action will have the terms in Eq.(\ref{actionabel}) traced over the gauge indexes, plus those known to appear in the flat space action, that come from ${\mathcal N}=4$. Up to dimension five operators\footnote{We need fermionic terms up to dimension five to get lepton masses, as we will see later.}, the fermionic non-abelian action is 
\bea \label{actionferm}
{\mathcal L}= Tr \left(-\frac{1}{2} \bar{\psi}^{\bi} \Gamma^\mu D_\mu \psi^i -\frac{1}{2} \bar{\lambda} \Gamma^\mu D_\mu \lambda + \frac{4\pi^2 \alpha'^2}{48} s_2 \psi^i \Phi^{\bi} \psi^j \Phi^{\bj} \right. \nonumber\\
 \left. + \frac{1}{48}G_0 \lambda \lambda + \frac{\pi^2 \alpha'^2}{12}G_2 \lambda \lambda \Phi^i \Phi^{\bi}  + \frac{1}{2 g_s} \epsilon_{ijk} \psi^i \psi^j \Phi^k + \frac{1}{2} [\psi^{\bi}, \Phi^i] \lambda \right) 
\eea
The first five terms are the ones that appear in the abelian action (\ref{actionabel}), applied to the background (\ref{fields}). The last two, that are zero in the abelian case, are those that come from the ${\mathcal N}=4$ theory. In particular, the next to the last term in Eq.(\ref{actionferm}) comes from the superpotential term $W=Tr \left([\Xi^i,\Xi^j] \Xi^k\right)$.
      
It remains to check that there are no more intrinsically non-abelian terms showing up in curved space, and we will do that with the help of the non-abelian D0-brane action found in \cite{Taylor2}. This action was found from matrix theory, and involves the fermionic world-volume coordinate $\Theta$, which is a 10D Majorana-Weyl fermion transforming in the adjoint of the gauge group. This fermion, when splitting the ${\bf \overline{16}}$ of $SO(9,1)$ 
into $({\bf\bar{2}},{\bf4})+({\bf 2},{\bf\bar{4}})$ of $SO(3,1)\otimes SO(6)$, gives the three fermions in the chiral multiplet $\psi^i$ and the gaugino $\lambda$.

The terms that we are interested in would involve a commutator of a scalar field $\Phi$ and the fermion $\Theta$ (and another power of $\Theta$), or the commutator of two $\Phi$'s together with a fermion bilinear.  From the results shown in \cite{Taylor2}, we can see that terms with a commutator of $\Phi$ and $\Theta$ appear in the 11-dimensional stress tensor $T^{ij}$, membrane currents $J^{ijk}$, $J^{+-i}$ and 5-brane current $M^{+-ijkl}$, where $+$ and $-$ are light cone coordinates, in the form
\bea
T^{ij}&\propto& {\overline \Theta} \Gamma^i [\Phi^j, \Theta] + {\overline \Theta} \Gamma^j [\Phi^i, \Theta] + ... \qquad M^{+-ijkl} \propto {\overline \Theta} \Gamma^{[ijk} [\Phi^{l]}, \Theta] +... 
\nonumber\\
J^{ijk} &\propto& {\overline \Theta} \Gamma^{0[ik} [\Phi^{k]}, \Theta] +... \qquad \quad \qquad \quad \quad \quad
 J^{+-i} \propto {\overline \Theta} \Gamma^0 [\Phi^i, \Theta] +... 
\eea
where ``$+...$'' indicates terms of a different structure. These currents couple to $h_{ij}$, $C_{ijk}$, $B_{0i}$ and $C_{0ijkl}$ in the IIA theory, as follows:
\be
{\cal L} \propto h_{ij} T^{ij}+ C_{ijk} J^{ijk} + 3 B_{0i} J^{+-i} + 6\, C_{0ijkl} M^{+-ijkl} +...
\ee

 Performing T-duality in 3 directions to get the D3-brane couplings, we get couplings to  $h_{ij}$, $h_{\mu\nu}$,  $C_{\mu i}$, $C_{\mu \nu  ij}$,$C_{\mu \nu \rho ijk}$, $B_{0i}$ and  $C_{\mu\nu\lambda\rho ijkl}$, where Greek letters indicate directions along the D3-brane. These are either 0 or second order in $\Phi$, giving higher order terms.

Terms of the form $[\Phi^i,\Phi^j] \Theta^2$ appear in the first moments of the currents, which couple to one derivative of the background fields (i.e. couple to the field strengths and not the potentials). Since  $[\Phi^i,\Phi^j] \Theta^2$ is already dimension 5, we need just to look at the possibility of a coupling of this form to $G_{ijk}$ or $G_{\bi \bj \bar{k}}$, the only constant field strengths. The T duals of these  couple to $J^{\pm ij(k)}$ and  $M^{-ij \mu\nu\lambda(k)}$, in the form
\be
{\cal L} \propto -2 \partial_k C_{ij \mu\nu\lambda} M^{-ij \mu \nu \lambda (k)} + \partial_k B_{ij} \left(J^{+ij(k)}-J^{-ij(k)}\right).
\ee 

 It can be checked from the original M-theory calculation done in \cite{TaylorM} that these moments of the currents do not contain such $[\Phi^i,\Phi^j] \Theta^2$ terms. So, for our background, there are not, up to dimension 5, any other intrinsically non-abelian terms than those already present in flat space, so the actions (\ref{actionbos}) and (\ref{actionferm}) build the full non-abelian action for a D3-brane in the background fluxes of Eq.(\ref{fields}).

A couple of comments regarding the action are in order: the quadratic and cubic terms in the bosonic part of the action (Eq.(\ref{actionbos})) break supersymmetry explicitly, and they come from fluxes that break supersymmetry in the bulk. On the contrary, the quartic piece in the bosonic action is SUSY preserving, as it comes from the superpotential $W=Tr \left([\Xi^i,\Xi^j] \Xi^k\right)$. In the fermionic action (Eq.(\ref{actionferm})), the terms proportional to $s_2$, $G_0$ and $G_2$ break supersymmetry (the first one because it cannot be of the form $\psi^i \psi^j \partial^2 W/ \partial_i \Phi \partial_j \Phi$, since $W$ cannot  depend on $\Phi^{\bi}$), and also come from supersymmetry breaking bulk supergravity fields. The last two terms preserve supersymmetry, the first one comes from the superpotential written above, and the last one from the Kahler part of the Lagrangian. 
The second comment is that fermions couple only to $S_{ij}, G_{ijk}$ and $A_{\bi \bj}$, as can be seen from Eq.(\ref{actionabel}). These pieces of the 3-form flux vanish in no-scale structure solutions \cite{NS}.

As far as the bosonic action, the scalar mass term also vanishes in no scale structure models, as we will see below.  The $(0,3)$ piece of the 3-form flux, which breaks supersymmetry but is present in no-scale structure solutions, appears in the bosonic action, but in a term that is intrinsically non-abelian, so it is not present in brane-world models with just one D-brane. The vanishing of tree level masses coming from supersymmetry breaking fluxes is a general prediction of no-scale structure models \cite{NSNM,DWG}. The cosmological constant vanishes for these models, even though supersymmetry is broken in the bulk.          

Now that we have the whole non-abelian action, we can write it in terms of the fields in irreducible representations, shown in Figure 1.
 Let us start with the bosonic part. Each term where an index $i$ is contracted with an $\bi$, like the mass term in
Eq.(\ref{actionbos}), gives bilinears of the form\footnote{All fields with (without) a tilde indicate the bosonic (fermionic) component of the corresponding supermultiplet in (\ref{matter}).} 
\be
Tr \left(\Phi^i \Phi^{\bi}\right) \rightarrow \sum_{lm}{\tilde{\cal Q}}^\dagger_l  {\tilde{\cal Q}}_m +  {\tilde{\cal L}}^\dagger_a  {\tilde{\cal L}}_a + {\tilde {\bar{\cal Q}}}^\dagger_a {\tilde{\bar {\cal Q}}}_a
\label{bilinear}
\ee
 Each term on the r.h.s. of this equation is multiplied  by a constant. These same constants appear also in the kinetic terms, so we can get rid of them by renormalizing the fields. 

A trilinear term of the form $\Phi^i \Phi^j \Phi^k$, as in the third term in Eq.(\ref{actionbos}), gives
\be
Tr \left(\Phi^i \Phi^j \Phi^k\right) \rightarrow  \sum_{la} \lambda_{la}  {\tilde{\cal Q}}_l   {\tilde{\cal L}}_a   {\tilde{\bar{\cal Q}}}_a
\ee
The way to understand this, and get the value of the coefficients $\lambda_{la}$, is the following: each $\Phi^i$ on the l.h.s. transforming originally in the adjoint 
of $U(27)$, corresponds, in the orbifolded theory, to a bifundamental $\phi^{a {\bar b}}$, a scalar field that goes from node $a$ to node $b$ in the quiver diagram of Figure 1. This identification has to be accompanied with a  Clebsh-Gordan coefficient  $Y^i_{a {\bar b}}$
coming from the projection $R_3\otimes R^a \rightarrow R^b$ as in Eq.(\ref{decomp}). A product of three $\Phi$'s then gives
\be \label{orbdecomp}
Tr \left(\Phi^i \Phi^j \Phi^k\right) \rightarrow 
Y^i_{a {\bar b}} Y^j_{b {\bar c}} Y^k_{c {\bar a}}\, \phi^{a {\bar b}} \phi^{b {\bar c}} \phi^{c {\bar a}}
\ee
So the trilinear coupling appearing in the third term of Eq.(\ref{actionbos}) is 
\be
Tr \left(G_{ijk} \Phi^i \Phi^j \Phi^k\right)  \rightarrow G_0 \sum_{la} \lambda_{la}  {\tilde{\cal Q}}_l   {\tilde{\cal L}}_a   {\tilde{\bar{\cal Q}}}_a
\label{trilinearA}
\ee
where
\be
\lambda_{la}= \epsilon_{ijk} Y^i_{(3_c {\bar3}_w)_l} Y^j_{(3_w {\bar 1}^a)} Y^k_{(1^a {\bar 3}_c)}
\label{CGtrilinearA}
\ee
and we have identified $a,b$ and $c$ in Eq.(\ref{orbdecomp}) with $U(3)_c, U(3)_w$ and $U(1)^a$ respectively.

The same can be done with the quartic terms, containing contractions $i \bi j \bj$, which will give terms of the form\footnote{The trace involves all possible loops with start and end points at a given node and made out of four arrows.} 
\bea
Tr \left(\Phi^j \Phi^{(i} \Phi^{\bi)} \Phi^{\bj} + h.c.\right) &&\rightarrow  \sum_{lma} \alpha_{lma} {\tilde{\cal Q}}_l {\tilde{\cal L}}_a {\tilde{\cal L}}^{\dagger}_a {\tilde{\cal Q}}^{\dagger}_m + 
\sum_{ab} \beta_{ab} {\tilde{\cal L}}_a {\tilde{\overline{\cal Q}}}_a {\tilde{\overline{\cal Q}}}^{\dagger}_b {\tilde{\cal L}}^{\dagger}_b +
\sum_{abl} \gamma_{alm} {\tilde{\overline{\cal Q}}}_a {\tilde{\cal Q}}_l {\tilde{\cal Q}}^{\dagger}_m {\tilde{\overline {\cal Q}}}^{\dagger}_a \nonumber\\
&+& \sum_{lmno} \eta_{lmno} {\tilde{\cal Q}}_l {\tilde{\cal Q}}^{\dagger}_m {\tilde{\cal Q}}_n {\tilde{\cal Q}}^{\dagger}_o + 
\sum_{ab} \rho_{ab} {\tilde{\cal L}}^{\dagger}_a {\tilde{\cal L}}_b {\tilde{\cal L}}^{\dagger}_b {\tilde{\cal L}}_a + 
\sum_{ab}  \nu_{ab} {\tilde{\overline{\cal Q}}}_a {\tilde{\overline{\cal Q}}}^{\dagger}_b {\tilde{\overline{\cal Q}}}_b {\tilde{\overline{\cal Q}}}^{\dagger}_a 
\label{quartic}
\end{eqnarray}
where, for example, $\alpha_{lma}$ can be obtained from the Clebsh-Gordan coefficients as   
\be
\alpha_{lma}= \sum_{ij} Y^j_{(3_c {\bar3}_w)_l} Y^i_{(3_c {\bar 1}^a)} Y^{\bi}_{(1^a {\bar3}_w)} Y^{\bj}_{(3_w {\bar3}_c)_m}
\ee
and equivalently for the rest of the coefficients.
After breaking $U(3)_w \times U(1)^3$ to $SU(2)_w \times U(1)_0$, we get all possible quartic terms involving either zero, two or four $SU(2)$ doublets contracted among them. 

\section{Masses and Couplings}
 
From the fermionic and bosonic actions (\ref{actionbos}) and (\ref{actionferm}), once decomposed into the fields that furnish irreducible representations of the orbifold group (Eqs.(\ref{bilinear}), (\ref{trilinearA}) and (\ref{quartic})), we get masses for all the scalars, trilinear A-terms, lepton Yukawa couplings and gaugino masses. Let us analyze them in detail, starting with the fermionic terms. 

The term  $\psi^i \psi^j \Phi^k$ in Eq.(\ref{actionferm}), independent of the fluxes, gives quark Yukawa couplings of the form
\be
Tr\left(\epsilon_{ijk}\psi^i\psi^j\Phi^k\right) \rightarrow \sum_{lm} \left( a_{lm} Q_l \tilde{H}_m {\bar u}_m +b_{lm} Q_l \tilde{\bar{H}}_m  \bar{d}_m\right)
\ee
where $a_{lm}$ and $b_{lm}$ can be obtained from the Clebsh-Gordan coefficients 
\be
a_{lm}= \sum_{ijk} \epsilon_{ijk}  Y^i_{(3_c {\bar2}_w)} Y^k_{(2_w {\bar 1}_{2,m})} Y^{j}_{(1_{2,m} {\bar3}_c)} 
\ee
and similarly for $b_{lm}$ (just change the node $1_{2,m}$ to $1_{3,m}$). We have taken into account the breaking of $U(3)_w \times U(1)^3$ into $SU(2)_w \times U(1)_0$.

Up and down quark masses are then given by
\be
(m_u)_{lm}=a_{lm}\langle H_m\rangle, \qquad (m_d)_{lm}=b_{lm}\langle \overline{H}_,\rangle 
\ee
If all $a_{lm}$ and $b_{lm}$ are the same, a hierarchy between generations can be obtained by having a hierarchy of VEVs.

It is worth noting that the symmetries of the orbifold forbid a contribution from the fluxes to quark masses. That contribution would appear 
if there was a linear term in $S_{ij}$, the symmetric combination of the $(1,2)$ piece of the 3-form flux (see Eq.(\ref{actionabel})),which is not compatible with the
orbifold projection.

As noted in \cite{Berenstein}, there are no lepton Yukawa couplings of this form, since the Higgs fields and the leptons belong to the same leg in the quiver of Figure 1.
Lepton masses come only at dimension five, from the term of the form $\psi^i \Phi^{\bi} \psi^j \Phi^{\bj}$ in Eq.(\ref{actionferm}). Projecting onto the fundamental matter of the orbifold, we get terms of the same form as those in (\ref{quartic}), with two scalars replaced by two fermions. The scalar fields that acquire a VEV are ${\tilde H}_m$, ${\tilde{\overline H}}_m$ and ${\tilde g}_m$, the first two are doublets of $SU(2)$ and the last one is a singlet. Then, lepton masses come from terms of the form
\be
 Tr \left( s_2\, \psi^i \Phi^{\bi} \psi^j \Phi^{\bj}\right) \rightarrow \sum_{lm} \left( s_2\rho_{lm} L_l {\tilde g}_l^{\dagger} {\overline e}_m {\tilde H}^{\dagger}_m + s_2 \rho'_{lm} L_l {\tilde g}_l^{\dagger} {\overline \nu}_m {\tilde{\overline H}}^{\dagger}_m \right)               
\ee
where
\be
\rho_{lm}= \sum_{ij} Y^i_{(2_w {\bar1}_{1,l})} Y^{\bi}_{(1_{1,l} {\bar 1}_0)} Y^{j}_{(1_0 {\bar1}_{1,m})} Y^{\bj}_{(1_{1,m} {\bar2}_w)}
\ee
and similarly for $\rho'$.  
Charged lepton masses and neutrino Dirac masses are then equal to
\be
(m_L)_{lm}= \frac{4 \pi^2 \alpha'^2}{48} s_2 \rho_{lm} \langle {\tilde g}^{\dagger}_l \rangle \langle {\tilde H}^{\dagger}_m \rangle, \qquad (m_D)_{lm}= \frac{4 \pi^2 \alpha'^2}{48} s_2 \rho'_{lm} \langle {\tilde g}^{\dagger}_l \rangle \langle {\tilde{\overline H}}^{\dagger}_m \rangle
\ee
 These masses involve the quadratic term in the $(1,2)$ piece of 3-form flux. As it was shown in \cite{GP,Gubser}, this flux breaks supersymmetry, and it is not of the no-scale form. Thus lepton masses in this model are supersymmetry (and ``no-scale structure'') breaking. Again, a hierarchy between generations can be obtained from a hierarchy of vevs
for ${\tilde H}_m$ and $ {\tilde{\overline H}}_m$. Neutrinos are lighter than charged leptons if $ \langle {\tilde{\overline H}}_m\rangle < \langle {\tilde H}_m \rangle$ (i.e. large $tan \beta$), 
which also gives lighter down than up quarks.       

From the same supersymmetry breaking 3-form flux we get also Higgsino masses. There is a generation mixing $\mu$-term, where $\mu$ is given by
\be
(\mu)_{lm}=\frac{4 \pi^2 \alpha'^2}{48} s_2 \rho''_{lm} \langle {\tilde H}^{\dagger}_l \rangle \langle {\tilde{\overline H}}^{\dagger}_m \rangle,
\ee
 
Finally, the $(3,0)$ piece of the 3-form flux, which also breaks supersymmetry, gives gaugino masses. All gauginos receive the same mass, proportional to $G_0$ to lowest order. 

If we turn on a gauge field, the fermionic Lagrangian computed in \cite{Grana} includes -non supersymmetric- electric and magnetic moment terms of the form
\begin{eqnarray*}
(2\pi \alpha') Tr \left(\epsilon_{ijk} \psi^i \gamma^{\mu \nu} \psi^j \,\partial_{\bar k} \tau \,(F + i \ast F)_{\mu \nu} \right) = (2\pi \alpha')^2\, \tau_2 \,Tr \left(\epsilon_{ijk} \psi^i \gamma^{\mu \nu} \psi^j \Phi^k  (F + i \ast F)_{\mu \nu}\right)
\end{eqnarray*}
where on the right hand side we have inserted the values for our background (Eq.(\ref{fields})).
This term breaks SUSY on the brane, as a nonvanishing antiholomorphic derivative of $\tau$ breaks SUSY in the bulk. 
Projecting onto the orbifolded matter, as in Eq.(\ref{trilinearA}), we get chromoelectric moments
\be
(2\pi \alpha')^2 \tau_2 \sum_{lm} (F + i\ast F)_{\mu\nu}\left(a_{lm} Q_l \gamma^{\mu \nu} \tilde{H}_m {\bar u}_m  + b_{lm} Q_l \gamma^{\mu\nu}  \tilde{{\overline H}}_m  \bar{d}_m \right) 
\ee
If $a_{lm}, b_{lm} \neq \delta_{lm}$, we get quark transition moments between different generations. For the same reason as there are no lepton Yukawa couplings with the Higgs field, we do not get lepton transition moments.

We turn our attention now to the bosons. From the second term in Eq.(\ref{actionbos}), projected as in Eq.(\ref{bilinear}), all scalar partners receive the same mass, given by 
\be
m^2= 4k_2+2 F_1\, g_s + 2 Im \tau_2\, g_s=\frac{g_s}{3} |G_0|^2 - \frac{2g_s}{3} G_0 G'_0 
\ee
where in the last equality we have used the conditions (\ref{F1det}-\ref{k2det}). This mass involves a priori the longitudinal metric, 5-form flux and second order dilaton, but not the 3-form flux, which appears only through the equations of motion. This combination vanishes  
for no-scale structure solutions, where $G_0=0$ \footnote{Even if we add D3-brane sources to the equations of motion and Bianchi identities, scalar masses would still be proportional to the (3,0) piece of the 3-form flux, as the D3-brane contribution to the metric and 5-form flux cancels out in the mass formula.}. 

From the supersymmetry breaking $(3,0)$ and $(0,3)$ pieces of the Lagrangian (third and fourth terms in Eq.(\ref{actionbos})), we get soft trilinear A-terms of the form
\bea
 G_0 \epsilon_{ijk}\Phi^i\Phi^j\Phi^k &+& G'_{0}\epsilon_{\bi\bj\bar{k}}\Phi^{\bi} 
\Phi^{\bj}\Phi^{\bar{k}} \rightarrow \nonumber\\
G_0 \left(\sum_{lm} a_{lm} {\tilde  Q}_l   {\tilde H}_m     {\tilde{\overline u}}_m + \sum_{lm} b_{lm} {\tilde  Q}_l   {\tilde{\overline H}}_m     {\tilde{\overline d}}_m \right) &+& G'_0 \left(\sum_{lm} a_{lm} {\tilde  Q}^{\dagger}_l   {\tilde H}^{\dagger}_m     {\tilde{\overline u}}^{\dagger}_m + \sum_{lm} b_{lm} 
{\tilde  Q}^{\dagger}_l   {\tilde{\overline H}}^{\dagger}_m     {\tilde{\overline d}}^{\dagger}_m \right)
\eea
One more time, the symmetries of the orbifold forbid soft terms of this type for the sleptons. 

The last term in (\ref{actionbos}), after projecting as in (\ref{quartic}), gives off-diagonal squark and slepton masses 
\be
(m^2_{\tilde Q})_{lm}= \sum_a \alpha_{lma} \langle g_a \rangle ^2, \qquad (m^2_{\tilde L})_{ab}= \mu_{ab} \langle g_a \rangle \langle g_b \rangle  
\ee
and diagonal Higgs boson masses
\be
(m^2_{\tilde H})_{b}= \sum_a \rho_{ab} \langle g_a \rangle ^2.
\ee

To analyze the order of magnitude of these masses and couplings, we assign a characteristic scale to $G_0$, the constant in the (3,0) piece of the 3-form flux. Since this component of the flux breaks supersymmetry, such scale should be identified with the supersymmetry breaking scale, denoted $m_{susy}$ in \cite{Berenstein}. Then, from the equations of motion, the parameters in the background fields of Eq.(\ref{fields}) have order of magnitude 
\be
G_0\sim m_{susy}\, , \quad G_2 \sim s_2 \sim m_{susy} M^2 \, ,\quad \tau_2 \sim m_{susy}^2
\ee
where $M$ is the string scale. 

Assuming all Clebsh-Gordan coefficients are of order one, we get quark masses of order $\langle H \rangle$. Lepton masses are suppressed by a factor $m_{susy}\langle g \rangle / M^2$ with respect to quark masses, as they come entirely from the 3-form flux. Gaugino masses are of order $m_{susy}$. The $\mu$-term is of order $m_{susy} \langle H \rangle^2/M^2$.  All scalars get diagonal masses of order $m_{susy}$, and squarks and sleptons get additional nondiagonal masses suppressed by a factor $\langle g \rangle / m_{susy}$ with respect to the diagonal ones. Trilinear A-terms are also of order $m_{susy}$. Quark transition moments are of order $m_{susy}^2 \langle H \rangle / M^4 \sim 10^{-11} \mu_B \langle H \rangle / M$.  

As suggested in \cite{Berenstein}, we get a semi-realistic spectrum  with $M \sim 10 TeV$, $\langle g \rangle \sim 1 TeV$, $m_{susy} \sim 3 TeV$,  although in this case higher order corrections are not much smaller than the terms we have considered. 
For these scales, Higgs masses are large, of the order of $1TeV$. Higgs VEVs are in the range $1MeV \sim 100 GeV$ to get a realistic quark spectrum, so dipole moments are considerably large, around the upper limits set by experiments.

Proton stability in ensured, since baryon number, being the $U(1)$ in $U(3)_c$ survives as a global symmetry after canceling the anomaly via a generalized Green-Schwarz mechanism.  

We have shown that by turning on background fluxes, we get in fact all masses and couplings predicted in \cite{Berenstein}. With our result, we can trace what flux is responsible for the different features in the Standard Model.

We should note that we expect the D-brane world-volume Lagrangian to receive corrections involving the blow-up parameters $r_a$ (using the notation in \cite{Berenstein}), controlled by twisted moduli. Since these are no more delocalized than D-brane fields, we should consider them in the action. But  we do not have a systematic way of computing these corrections, which involve in principle string world-sheet computations. Nevertheless, we should expect them to be of order $r_a \alpha' \sim \langle g \rangle^2/M^2$ which is small in the scenario presented in \cite{Berenstein}. 

Twisted moduli will also change the gauge couplings for each sector \cite{Lawrence} from Eq.(\ref{couplings}) to 
\be
\tau_i= \int_{\Sigma_i} C_{(2)} + \tau  \int_{\Sigma_i} B_{(2)}
\ee
where $\Sigma_i$ are the 2-cycles associated with each conjugacy class.  It should be possible to tune these moduli to have the desired $\theta_w$ angle.

\section{Conclusions}

We have obtained 
the parameters of the MSSM that one gets when placing a D3-brane at a $\Delta_{27}$ orbifold singularity
with susy breaking background fluxes turned on, as a function of these fluxes. We could trace gaugino and Higgsino masses, scalar masses,
trilinear couplings and dipole moments as
the effects of supersymmetry breaking in the bulk. 

Semirealistic spectra can be obtained from breaking SUSY in the bulk at a TeV scale.
In order to get realistic lepton masses, the string scale needs to be considerably low, just an order of magnitude larger
than the SUSY breaking scale. 
The model has a rich phenomenology that can be studied in terms of different supergravity
backgrounds.  

In this paper we have only considered backgrounds that preserve Lorentz invariance on the world-volume.
We can consider a more general case, where for example the longitudinal components of the metric can fluctuate,
and obtain the parameters of softly broken ${\mathcal N}=1$ supergravity models in terms of background fields.    

This paper opens a bridge between MSSM phenomenology and compactifications in background fluxes. 
The model of Berenstein, Jejjala and Leigh has the advantage of simplicity, as it consists of just one D3-brane,
but similar calculations can in principle be performed for other models found in the literature, opening up new 
avenues to study their phenomenology.

\section*{Acknowledgements}
I am deeply indebted to Joe Polchinski for guidance through the project. 
I would also like to thank Gerardo Aldazabal, Vishnu Jejjala and Robert Leigh for useful discussions.
This work was supported by National Science Foundation grant PHY97-22022. 

\appendix
\section{The group $\Delta_{27}$}
In this Appendix we show the basic features of the group $\Delta_{27}$, and 
show how to get the number of fields transforming in the $(N_a, {\overline N_b})$ 
representation.

The group $\Delta_{27}$ is one of the non abelian subgroups of $SU(3)$, of 
the form $\Delta_{3n^2}$ (see for example \cite{Fairbairn, Muto}). It is 
given by the following elements (this is the defining representation of 
$\Delta _{27}$)
\bea
A_{i,j}=\pmatrix{w^i & 0 & 0 \cr
0 & w^j & 0 \cr
0 & 0 & w^{-i-j} } \quad
C_{i,j}=\pmatrix{0 & 0 & w^i \cr
w^j & 0 & 0 \cr
0 & w^{-i-j} & 0 }\quad
E_{i,j}=\pmatrix{0 & w^i & 0 \cr
0 & 0 & w^j \cr
w^{-i-j} & 0 & 0 } \quad
\label{elements}
\eea
where $w=e^{\frac{2}{3}\pi i}$ and $0 \le i,j < 3$. So there are 27 elements 
in the group. These can be grouped in 11 conjugacy classes (there are as 
many conjugacy classes as irreducible representations) as \footnote{The 
notation follows that used in the literature, which allows the 
characterization of all $\Delta_{3n^2}$ groups.}
\bea
C_1^0&=&\{A_{0,0}\}\hspace{0.9cm}  \qquad  \quad C_1^1=\{A_{1,1}\} \hspace{0.9cm}  \qquad  \quad  C_1^2=\{A_{2,2}\} \nonumber\\
 C_2^{12}&=&\{A_{0,1}; A_{2,0};A_{1,2}\} \; \;  C_2^{21}=\{A_{1,0}; A_{0,2}; 
A_{2,1}\} \nonumber\\
C_3^0&=&\{C_{0,0}; C_{1,1}; C_{2,2}\} \quad  C_3^1=\{C_{1,0}; C_{2,1}; C_{0,2}\} \quad
 C_3^2=\{C_{2,0}; C_{0,1}; C_{1,2}\} \nonumber\\
C_4^0&=&\{E_{0,0}; E_{1,1}; E_{2,2}\} \quad 
C_4^1=\{E_{1,0}; E_{2,1}; E_{0,2}\} \quad   
C_4^2=\{E_{2,0}; E_{0,1}; E_{1,2}\}  \nonumber
\eea

There are nine 1-dimensional irreducible representations for this group, 
called $R_1^a$ ($R_1^0$ is the trivial representation), and two 
3-dimensional irreducible representations, $R_3^1$ and $R_3^2$, where  
$R_3^1$ is the defining representation shown above.
In the Table I, we show the characters of the conjugacy classes for 
all of these representations.

\begin{table}[t]
\begin{center}
\begin{tabular}{|c|c|c|c|c|}
\hline
& $C_1^l $  & $C_2^{i,j}$ & $C_3^l$ & $C_4^l$ \\
\hline
$|C_a|$ &  1 & 3 & 3  & 3  \\
\# classes & 3 & 2 & 3 & 3\\
$R_1^0, R_1^1, R_1^2$ & 1 & 1 & $1,w,w^2$ & $1,w^2,w$\\
$R_1^3, R_1^4, R_1^5$ & 1 & $w^{i-j}$ & $w^l, w^{1+l}, w^{2+l}$ &  $w^l, 
w^{2+l}, w^{1+l}$  \\
$R_1^6, R_1^7, R_1^8$ &  1 & $w^{j-i}$ & $w^{2l}, w^{1+2l}, w^{2+2l}$ &  
$w^{2l}, w^{2+2l}, w^{1+2l}$  \\
$R_3^1$ & $3w^l$ & $w^i+w^{-(i+j)}+w^j$ & 0 & 0 \\
$R_3^2$ & $ 3w^{2l}$ & $w^{i+j}+w^{-j}+w^{-i}$ & 0 & 0  \\

\hline
\end{tabular}
\caption{Character table for the group $\Delta_{27}$}
\end{center}
\end{table}

The number of fields transforming as $(N_a,{\overline N_b})$ comes from the 
decomposition of the product of the defining and each irreducible representation 
of a group $G$ into irreducible representations
\be
R_3 \otimes R^a = \oplus_{b=1}^r n^3_{ab} R^b.
\label{decomp}
\ee
To obtain it, we note that any reducible representation can be decomposed 
into a sum of irreducible representations
\be
R = \oplus_{a=1}^r n_a R^a
\ee
where $n_a$ is the number of times the representation $R^a$ appears in $R$ 
(for $R=R_{reg}$, $n_a=dim R^a$). This means that for any element $g$ of the 
group $G$, we can get its character $\chi^R(g)$ in the representation $R$ by 
summing over its characters in each irreducible representation as
\be
\chi^R(g)= \sum_{a=1}^r n_a \chi^a(g).
\ee
From this, we can obtain $n_a$ as
\be
n_a=\frac{1}{|G|}\sum_{g \in G} \chi^R(g) \chi^a(g)^*
\label{na}
\ee
where we have used the orthogonality condition
\be
\frac{1}{|G|}\sum_{g \in G} \chi^a(g) \chi^b(g)^*=\delta_{ab}.
\ee

In the case of a product of representations
\be
R^i\otimes R^j=R^k
\ee
the character of each element is also a product
\be
\chi^i(g) \chi^j(g)=\chi^k(g).
\label{prodchi}
\ee
Then, with $i$ being the defining representation and $j$ one of the 
irreducible representations, using Eqs.(\ref{na}) and (\ref{prodchi}) we can get 
the desired coefficients $n^3_{ab}$ as
\be
n^3_{ab}=\frac{1}{|G|}\sum_{g \in G} \chi^3(g)\chi^a(g) \chi^b(g)^*.
\ee
Since the elements of a group can be classified into conjugacy classes, and 
all the elements in a conjugacy class have the same character, we can 
rewrite the sum as a sum of conjugacy classes as
\be
n^3_{ab}=\frac{1}{|G|}\sum_{c=1}^r |C_c|\chi^3(C_c)\chi^a(C_c) \chi^b(C_c)^*
\ee

Then, using the table above, we can obtain all the numbers $n^3_{ab}$ (the 
characters $\chi^3(C_c)$ can be obtained from the row corresponding to 
$R_3^1$, since this irreducible representation is the defining 
representation). This gives
\be
n^3_{R_3^1, R_3^2}=3, \qquad n^3_{R_3^2, R_1^a}=1, \qquad 
n^3_{R_1^a,R_3^1}=1
\ee
for each $a$, and the rest of the $n^3_{R^a,R^b}$ are zero. This is the 
amount of matter claimed in \cite{Berenstein}, where the $R_3^1$ is 
identified with color, $R_3^2$ corresponds to the weak interaction, the 3 
multiplets in $n^3_{R_3^1,\bar{R_3^2}}$ are the ${\cal Q}_i$'s, each 
multiplet in  $n^3_{R_3^2, R_1^a}$ is called ${\overline {\cal Q}}_a$, and the 
ones in $n^3_{R_1^a,R_3^1}$ are the ${\cal L}_a$'s.

\end{document}